\documentclass[11pt]{article}

\usepackage{graphics}

\def\leptontag{{\tt Lepton}}
\def\kaonitag{{\tt Kaon\,I}}
\def\kaoniitag{{\tt Kaon\,II}}
\def\kpitag{{\tt Kaon-Pion}}
\def\piontag{{\tt Pion}}
\def\othertag{{\tt Other}}
\def\nbb{(347.5 \pm 3.8) \times 10^6\, \FourS \to B\Bbar}
\def\extranbb{120 \times 10^6\, \FourS \to B\Bbar}

\def\measures2twob{\fitstwob \pm \statstwob \pm \syststwob \syst}
\def\fits2twobstat{\fitstwob \pm \statstwob}

\def\measurerperp{\fitr \pm \statr \stat \pm \systr \syst}
\def\effectiveeta{0.504 \pm 0.033}

\def\fitr{0.233}
\def\statr{0.010}
\def\systr{0.005}
\def\fitstwob{0.710}
\def\statstwob{0.034}
\def\syststwob{0.019}
\def\fitlambda{0.932}
\def\statlambda{0.026}
\def\systlambda{0.017}

\newcommand{\bflav}{\ensuremath{\B_{{\rm flav}}}}

\newcommand{\BABARPubYear}    {06}

\newcommand{\BABARConfNumber} {036}
\newcommand{\SLACPubNumber} {11986}

\RequirePackage{xspace}

\usepackage{relsize}
\def\babar{\mbox{\slshape B\kern-0.1em{\smaller A}\kern-0.1em
    B\kern-0.1em{\smaller A\kern-0.2em R}}}

\def\epem       {\ensuremath{e^+e^-}\xspace}

\def\ccbar {\ensuremath{c\overline c}\xspace}

\def\piz   {\ensuremath{\pi^0}\xspace}

\def\pim   {\ensuremath{\pi^-}\xspace}

\def\Kbar  {\kern 0.2em\overline{\kern -0.2em K}{}\xspace}

\def\Kz    {\ensuremath{K^0}\xspace}
\def\Kzb   {\ensuremath{\Kbar^0}\xspace}
\def\KzKzb {\ensuremath{\Kz \kern -0.16em \Kzb}\xspace}
\def\Kp    {\ensuremath{K^+}\xspace}
\def\Km    {\ensuremath{K^-}\xspace}
\def\Kpm   {\ensuremath{K^\pm}\xspace}

\def\KpKm  {\ensuremath{\Kp \kern -0.16em \Km}\xspace}
\def\KS    {\ensuremath{K^0_{\scriptscriptstyle S}}\xspace} 
\def\KL    {\ensuremath{K^0_{\scriptscriptstyle L}}\xspace} 
\def\Kstarz  {\ensuremath{K^{*0}}\xspace}

\def\Dbar    {\kern 0.2em\overline{\kern -0.2em D}{}\xspace}

\def\Dz      {\ensuremath{D^0}\xspace}
\def\Dzb     {\ensuremath{\Dbar^0}\xspace}
\def\DzDzb   {\ensuremath{\Dz {\kern -0.16em \Dzb}}\xspace}
\def\Dp      {\ensuremath{D^+}\xspace}
\def\Dm      {\ensuremath{D^-}\xspace}

\def\DpDm    {\ensuremath{\Dp {\kern -0.16em \Dm}}\xspace}

\def\B       {\ensuremath{B}\xspace}
\def\Bbar    {\kern 0.18em\overline{\kern -0.18em B}{}\xspace}

\def\Bz      {\ensuremath{B^0}\xspace}
\def\Bzb     {\ensuremath{\Bbar^0}\xspace}
\def\BzBzb   {\ensuremath{\Bz {\kern -0.16em \Bzb}}\xspace}
\def\Bu      {\ensuremath{B^+}\xspace}
\def\Bub     {\ensuremath{B^-}\xspace}

\def\Bpm     {\ensuremath{B^\pm}\xspace}

\def\BpBm    {\ensuremath{\Bu {\kern -0.16em \Bub}}\xspace}

\def\BorBbar    {\kern 0.18em\optbar{\kern -0.18em B}{}\xspace}
\def\DorDbar    {\kern 0.18em\optbar{\kern -0.18em D}{}\xspace}
\def\KorKbar    {\kern 0.18em\optbar{\kern -0.18em K}{}\xspace}

\def\jpsi     {\ensuremath{{J\mskip -3mu/\mskip -2mu\psi\mskip 2mu}}\xspace}
\def\psitwos  {\ensuremath{\psi{(2S)}}\xspace}

\def\etac     {\ensuremath{\eta_c}\xspace}

\def\chicone  {\ensuremath{\chi_{c1}}\xspace}

\mathchardef\Upsilon="7107
\def\Y#1S{\ensuremath{\Upsilon{(#1S)}}\xspace}% no space before {...}!

\def\FourS {\Y4S}

\mathchardef\Deltares="7101
\mathchardef\Xi="7104
\mathchardef\Lambda="7103
\mathchardef\Sigma="7106
\mathchardef\Omega="710A

\def\Deltabar{\kern 0.25em\overline{\kern -0.25em \Deltares}{}\xspace}
\def\Lbar{\kern 0.2em\overline{\kern -0.2em\Lambda\kern 0.05em}\kern-0.05em{}\xspace}
\def\Sigbar{\kern 0.2em\overline{\kern -0.2em \Sigma}{}\xspace}
\def\Xibar{\kern 0.2em\overline{\kern -0.2em \Xi}{}\xspace}
\def\Obar{\kern 0.2em\overline{\kern -0.2em \Omega}{}\xspace}
\def\Nbar{\kern 0.2em\overline{\kern -0.2em N}{}\xspace}
\def\Xb{\kern 0.2em\overline{\kern -0.2em X}{}\xspace}

\def\mes        {\mbox{$m_{\rm ES}$}\xspace}

\newcommand{\tev}{\ensuremath{\mathrm{\,Te\kern -0.1em V}}\xspace}
\newcommand{\gev}{\ensuremath{\mathrm{\,Ge\kern -0.1em V}}\xspace}
\newcommand{\mev}{\ensuremath{\mathrm{\,Me\kern -0.1em V}}\xspace}
\newcommand{\kev}{\ensuremath{\mathrm{\,ke\kern -0.1em V}}\xspace}
\newcommand{\ev}{\ensuremath{\mathrm{\,e\kern -0.1em V}}\xspace}
\newcommand{\gevc}{\ensuremath{{\mathrm{\,Ge\kern -0.1em V\!/}c}}\xspace}
\newcommand{\mevc}{\ensuremath{{\mathrm{\,Me\kern -0.1em V\!/}c}}\xspace}
\newcommand{\gevcc}{\ensuremath{{\mathrm{\,Ge\kern -0.1em V\!/}c^2}}\xspace}
\newcommand{\mevcc}{\ensuremath{{\mathrm{\,Me\kern -0.1em V\!/}c^2}}\xspace}
\def\invfb   {\ensuremath{\mbox{\,fb}^{-1}}\xspace}

\def\mus  {\ensuremath{\rm \,\mus}\xspace}

\def\ps   {\ensuremath{\rm \,ps}\xspace}

\def\mus        {\ensuremath{\,\mu{\rm s}}\xspace}    %% microsecond
\def\ps         {\ensuremath{{\rm \,ps}}\xspace}  %% picosecond

\def\to                 {\ensuremath{\rightarrow}\xspace}

\newcommand{\stat}{\ensuremath{\mathrm{(stat)}}\xspace}
\newcommand{\syst}{\ensuremath{\mathrm{(syst)}}\xspace}

\def\pep2{PEP-II}

\def\gsim{{~\raise.15em\hbox{$>$}\kern-.85em
          \lower.35em\hbox{$\sim$}~}\xspace}
\def\lsim{{~\raise.15em\hbox{$<$}\kern-.85em
          \lower.35em\hbox{$\sim$}~}\xspace}

\def\eps{\varepsilon\xspace}

\def\CP                {\ensuremath{C\!P}\xspace}
\def\stwob{\ensuremath{\sin\! 2 \beta   }\xspace}

\def\mistag{\ensuremath{w}\xspace}

\def\deltaz{\ensuremath{{\rm \Delta}z}\xspace}
\def\deltat{\ensuremath{{\rm \Delta}t}\xspace}
\def\deltamd{\ensuremath{{\rm \Delta}m_d}\xspace}

\xspace

\newcommand{\jprlBase}       {Phys.\ Rev.\ Lett.\xspace}
\newcommand{\jprBase}        {Phys.\ Rev.\xspace}

\newcommand{\nimBaseA}       {Nucl.\ Instr.\ Methods Phys.\ Res., Sect.\ A\xspace}

\newcommand{\npBase}         {Nucl.\ Phys.\xspace}

\newcommand{\jpg}       [1]  {{J.\ Phys.\ {\bf G{\bf #1}}}}

\newcommand{\nima}      [1]  {\nimBaseA~{\bf #1}}

\newcommand{\np}        [1]  {\npBase\ {\bf #1}}
\newcommand{\jprl}      [1]  {\jprlBase\ {\bf #1}}
\newcommand{\jprd}      [1]  {\jprBase\ D~{\bf #1}}
\def\jetset74   {\mbox{\tt Jetset \hspace{-0.5em}7.\hspace{-0.2em}4}\xspace}
\long\def\inst#1{\par\nobreak\kern 4pt\nobreak
    {\it #1}\par\vskip 10pt plus 3pt minus 3pt}

\begin{document}
{\pagestyle{empty}

\begin{flushright}
%BAD 1548 Version 9\\
\babar-CONF-\BABARPubYear/\BABARConfNumber \\
%\babar-PUB-\BABARPubYear/\BABARPubNumber \\
SLAC-PUB-\SLACPubNumber \\
%hep-ex/\LANLNumber \\
\end{flushright}

\par\vskip 5cm

% Title of the paper
\begin{center}
\Large \bf Improved Measurement of {\boldmath\CP} Asymmetries in 
{\boldmath $\Bz \to (\ccbar) K^{(*)0}$} Decays
\end{center}
\bigskip

\begin{center}
\large The \babar\ Collaboration\\
\mbox{ }\\
\today
\end{center}
\bigskip \bigskip

%---------------------------------
% Abstract
\begin{center}
\large \bf Abstract
\end{center}
%---------------------------------
We present an updated measurements of time-dependent \CP
asymmetries in fully reconstructed neutral $B$ decays to several
\CP eigenstates containing a charmonium meson.
The measurements use a data sample of 
$\nbb$ decays collected
with the \babar\  detector at the
PEP-II $B$ factory between 1999 and 2006.
We determine  $\stwob = \fitstwob \pm \statstwob \stat \pm \syststwob\syst$
and $|\lambda| = \fitlambda \pm \statlambda \stat \pm \systlambda \syst$.
Both of these results are preliminary.
\vfill
\begin{center}

Submitted to the 33$^{\rm rd}$ International Conference on High-Energy Physics, ICHEP 06,\\
26 July---2 August 2006, Moscow, Russia.

\end{center}

\vspace{1.0cm}
\begin{center}
{\em Stanford Linear Accelerator Center, Stanford University, 
Stanford, CA 94309} \\ \vspace{0.1cm}\hrule\vspace{0.1cm}
Work supported in part by Department of Energy contract DE-AC03-76SF00515.
\end{center}

\newpage
} % end of pagestyle{empty}

% Input author list file
%
%author list removed temporarily to save trees 7/9/04 RNC
%\input authors_ICHEP2006_bad1548.tex
\begin{center}
\small

The \babar\ Collaboration,
\bigskip

%% author list as of 01-Jul-2006 (596 authors)
%
{B.~Aubert,}
{R.~Barate,}
{M.~Bona,}
{D.~Boutigny,}
{F.~Couderc,}
{Y.~Karyotakis,}
{J.~P.~Lees,}
{V.~Poireau,}
{V.~Tisserand,}
{A.~Zghiche}
\inst{Laboratoire de Physique des Particules, IN2P3/CNRS et Universit\'e de Savoie,
 F-74941 Annecy-Le-Vieux, France }
% ...

\end{center}\newpage

% The body of the paper starts here
%------------------------------
\section{INTRODUCTION}
\label{sec:Introduction}
%------------------------------
Charge conjugation-parity (\CP) violation in the $B$ meson system has been
established by the \babar~\cite{babar-stwob-prl}
and Belle~\cite{belle-stwob-prl} collaborations.
The Standard Model (SM) of electroweak interactions describes \CP\ violation
as a consequence of a complex phase in the
three-generation Cabibbo-Kobayashi-Maskawa (CKM) quark-mixing
matrix~\cite{ref:CKM}. In this framework, measurements of \CP\ asymmetries in
the proper-time distribution of neutral $B$ decays to
\CP\ eigenstates containing a charmonium and $K^{0}$ meson provide
a direct measurement of $\stwob$~\cite{BCP}. The angle $\beta$ is defined
as $\arg [-(V_{\rm cd}^{}V_{\rm cb}^*) / (V_{\rm td}^{}V_{\rm tb}^*)]$, where
the $V_{ij}$ are CKM matrix elements. 

In this paper, we report on an updated measurement
of $\stwob$ in $\nbb$ decays collected with the \babar\ detector using 
$B^0$ decays to the final states $\jpsi\KS$, $\jpsi\KL$, $\psitwos\KS$, $\chicone\KS$, $\eta_c \KS$, 
and $\jpsi\Kstarz (\Kstarz \to \KS\piz)$.
Since our previous measurement~\cite{ref:babar2004}, we have added a sample of 
$\extranbb$ decays, and applied an improved event reconstruction to the complete dataset.
A new $\eta_c \KS$ event selection has been
developed based on the Dalitz structure of the $\etac \to \KS \Kp \pim$ decay. We have also performed 
a more detailed study of the \CP\ properties of our background events resulting in a reduced
systematic error.
%We limit the discussion to the changes in the analysis with respect to the most recent
%published results in Ref.~\cite{ref:babar2004}.

%-----------------------------------------------------------------------------
\section{THE \babar\ DETECTOR AND DATASET}
\label{sec:babar}
%-----------------------------------------------------------------------------
The data used in this analysis were collected with the \babar\ detector
at the \pep2\ asymmetric $\epem$ storage ring from 1999 to 2006. This
represents a total integrated luminosity of ($316.2 \pm 3.5$) $\invfb$ taken at the
$\FourS$ resonance (onpeak), corresponding to a sample of $\nbb$
 decays.

The \babar\ detector is described in detail elsewhere~\cite{ref:babar}.
Charged particles are selected and their momenta are measured by a combination
of a vertex tracker consisting of five layers of double-sided silicon microstrip
detectors and a 40-layer central drift chamber, both operating in the 1.5~T magnetic
field of a superconducting solenoid. 
We identify photons and electrons using a CsI(Tl) electromagnetic calorimeter (EMC).
Charged particle identification (PID) is provided by an internally reflected 
ring imaging Cherenkov detector (DIRC) covering the central region of the
detector, the average energy loss ($dE/dx$) in the tracking devices, and by the EMC.
In addition, the instrumented flux return (IFR) containing resistive plate chambers
are used for muon and long-lived neutral hadron identifications. 
We use the GEANT4~\cite{ref:geant} software
to simulate interactions of particles traversing the \babar\ detector.

%------------------------------
\section{ANALYSIS METHOD}
\label{sec:Analysis}
%------------------------------
We use information
from the other $B$ meson, $B_{\rm tag}$, in the event to tag the initial flavor
of the fully reconstructed  $B$ candidate.
The decay rate $f_+ (f_-)$ for a neutral $B$ meson decaying to a \CP\ eigenstate 
accompanied by  a $B^{0}
(\Bzb)$ tag can be expressed in terms of a complex parameter 
$\lambda$~\cite{ref:lambda} as
\begin{equation}
{f}_\pm(\, \deltat) = {\frac{e^{{- \left| \deltat \right|}/\tau_{\Bz} }}{4\tau_{\Bz} }}
\Bigg\{ (1\mp\Delta\omega) \Bigg.  \pm  (1-2\omega) \times
\Big[{\frac{2\mathop{\cal I\mkern -2.0mu\mit m}\lambda}{1 + |\lambda|^2} }
  \sin{( \Delta m_{d}  \deltat )} -
 \Bigg. {\frac{1  - |\lambda|^2 } {1 + |\lambda|^2} }
       \cos{( \Delta m_{d}  \deltat) }  \Big] \Bigg\},
\label{eq:timedist}
\end{equation}
where $\Delta t = t_{\rm rec} - t_{\rm tag}$ 
is the difference between the proper decay times of the reconstructed ($B_{\rm rec}$) 
and tagged ($B_{\rm tag}$) $B$ mesons.
$\tau_{\Bz}$ is the \Bz lifetime and \deltamd is the mass difference determined from 
\Bz-\Bzb oscillations~\cite{ref:pdg2006}. 
The average mistag probability $\omega$ describes the effect of
incorrect tags, and $\Delta\omega$ is the difference between
the mistag rate for $\Bz$ and \Bzb .
Here we assume that the decay width difference $\Delta \Gamma_d$ between the neutral 
\B\ mass eigenstates is zero.
The sine term in Eq.~\ref{eq:timedist} is due to the interference between the 
direct decay and the decay after \Bz-\Bzb oscillation. A non-zero cosine term 
arises from the interference between decay amplitudes with different weak
and strong phases (direct \CP\ violation) or from \CP\ violation in
\Bz-\Bzb mixing.

In the SM, \CP\ violation in mixing is negligible, as is direct \CP
violation for $b \to \ccbar s$ decays that contain a charmonium meson~\cite{ref:lambda}.
Under these assumptions, $\lambda=\eta_f e^{-2i\beta}$ where $\eta_f=\pm 1$ is the \CP
eigenvalue of the final state $f$. Thus, the time-dependent 
\CP-violating asymmetry is
\begin{eqnarray}
A_{\CP}(\deltat) &\equiv&  \frac{ {f}_+(\deltat)  -  { f}_-(\deltat) }
{ {f}_+(\deltat) + {f}_-(\deltat) } \propto -(1-2\omega)\eta_f \stwob
\sin{ (\deltamd \, \deltat )}.
\label{eq:asymmetry}
\end{eqnarray}
%\vskip12pt\noindent

We reconstruct a sample of neutral $B$ mesons, $B_{\CP}$, decaying to the $\eta_f=-1$ final states~\cite{ref:chargeconj} 
$\jpsi\KS$, $\psitwos\KS$, $\chicone\KS$ and $\eta_c \KS$, and the $\eta_f=+1$ final 
state $\jpsi\KL$. The $\jpsi$ and $\psitwos$ mesons are reconstructed through their decays
to $e^+e^-$ and $\mu^+\mu^-$; the $\psitwos$ is also reconstructed through its decay to $\jpsi\pi^+\pi^-$.
The $\chicone$ meson is reconstructed in the decay mode $\jpsi\gamma$. The $\eta_c$ meson
is reconstructed in the decay mode $\KS\Kp\pim$.
We also reconstruct the $\jpsi\Kstarz (\Kstarz \to \KS\piz)$ 
final state which can be \CP -even or \CP -odd 
due to the presence of even ($L$=0, 2) and odd ($L$=1) orbital angular momenta contributions.
Ignoring the angular information in $\jpsi\Kstarz$ results in a reduction of the measured \CP\
asymmetry by a factor $\vert 1-2R_{\perp} \vert$, where
$R_{\perp}$ is the fraction of the $L$=1 contribution. We have measured
$R_{\perp} = \measurerperp$~\cite{ref:rperp}, which 
gives an effective $\eta_f = \effectiveeta$, after acceptance corrections.

In addition to the \CP\ modes described above, a large sample \bflav\ of \Bz\ decays to the flavor
eigenstates $D^{(*)-}h^+ (h^+=\pi^+,\rho^+$, and $a_1^+)$ and $\jpsi\Kstarz
(\Kstarz\to\Kp\pim)$ is used for calibrating the flavor tagging performance 
and \deltat resolution.
We perform studies to measure apparent \CP\ violation from unphysical sources using a 
control sample of $B^+$ mesons decaying to the final states $\jpsi K^{(*)+}$, $\psitwos
K^+$, $\chicone \Kp$, and $\eta_c \Kp$.
Since the previous \babar\ analyses, 
we apply an improved event reconstruction to all the events and we find that on a 
subset of the data the new and  old event reconstruction algorithms 
give consistent results.
The event selection and candidate reconstruction are unchanged from those described in 
Refs.~\cite{ref:babar2004,ref:bigprd,ref:etacks} with the 
exceptions described below. As in Ref.~\cite{ref:babar2004} we reconstruct
the $\Bz \to \etac \KS$ and $\Bpm \to \etac \Kpm$ modes using only the $\etac \to \KS \Kp \pim$ 
decay with $2.91 < m_{\KS \Kp \pim} < 3.05 \gevcc$.
 We now exploit the fact that the $\etac$ decays
predominantly through a $K \pi$ resonance at around $1430\mevcc$, and apply the 
selection criteria  
$1.26\gevcc < m(\KS\pim) < 1.63\gevcc$ and $1.26\gevcc < m(\Kp\pim) < 1.63\gevcc$.

We calculate the time interval \deltat between the two $B$ decays from the measured separation 
\deltaz between the decay vertices of $B_{\rm rec}$ and $B_{\rm tag}$ along the collision 
($z$) axis~\cite{ref:bigprd}. The $z$ position of the $B_{\rm rec}$ vertex is determined from
the charged daughter tracks. The $B_{\rm tag}$ decay vertex is determined by fitting tracks 
not belonging to the $B_{\rm rec}$ 
candidate to a common vertex, employing constraints from the beam spot
location and the $B_{\rm rec}$ momentum~\cite{ref:bigprd}. 
Events are accepted if the calculated \deltat\ uncertainty is less than 2.5\ps
and $\vert \deltat \vert$ is less than  $20\,\ps$. The fraction of events satisfying 
these requirements is 95\%.
The r.m.s. \deltat resolution is 1.1\ps for the 99.7\% of events that 
are not outliers.

Multivariate algorithms are used to identify signatures of $B$ decays that determine 
(``tag'') the flavor of the $B_{\rm tag}$ at 
decay to be either a \Bz\ or \Bzb candidate. 
Primary leptons from semileptonic $B$ decays are selected 
from identified electrons and muons as well as isolated energetic tracks. 
The charges of identified  kaon candidates are used in a kaon tag. 
Low momentum pions from $D^{*+}$ decays 
are selected on the basis 
of their momentum and direction with respect to the thrust axis of $B_{\rm tag}$. These algorithms 
are combined to account for correlations among different sources of flavor information and to 
provide an estimate of the mistag probability for each event. 
Each event whose estimated mistag probability is less than 45\% is assigned to one of six 
tagging categories.
The \leptontag\ category contains events with an identified lepton;
the remaining events are divided into the 
\kaonitag, \kaoniitag, \kpitag, \piontag, or \othertag\ categories based on
the estimated mistag probability.  
For each category $i$, the tagging efficiency $\eps_i$ and 
fraction $\mistag_i$ of events having the wrong tag assignment are measured from data 
(Table~\ref{tab:mistag}).
The figure of merit for tagging is the effective tagging efficiency
$Q \equiv \sum_i {\eps_i (1-2\mistag_i)^2} = (30.4 \pm 0.3)\,\%$, where the
error shown is statistical only.
%~\cite{ref:babar2004}.
%-----------------------------
% Table : Tagging
%-----------------------------
\begin{table}[!ht]
\begin{center}
\vspace{0.5cm}
\caption{Efficiencies $\eps_i$, average mistag fractions $\mistag_i$, mistag fraction differences
$\Delta\mistag_i \equiv \mistag_i(\Bz)-\mistag_i(\Bzb)$, and effective tagging 
efficiency $Q_i$ extracted for each tagging
category $i$ from the $B_{\rm flav}$ sample.\label{tab:mistag}}
\vspace{0.5cm}
\begin{tabular}{lcccc}\hline\hline
Category     &
{$\varepsilon$   (\%)} &
{$\mistag$       (\%)} &
{$\Delta\mistag$ (\%)} &
{$\ \ \ Q$             (\%)} \\ \hline
  \leptontag &   $\phantom{0}8.67\pm0.08$  &   $\phantom{1}3.0\pm0.3$  &  $-0.2\pm0.6$  & $\phantom{0}7.67\pm0.13$     \\
   \kaonitag &   $10.96\pm0.09$ &   $\phantom{0}5.3\pm0.4$  &  $-0.6\pm0.7$ & $\phantom{0}8.74\pm0.16$    \\
  \kaoniitag &   $17.21\pm0.11$ &   $15.5\pm0.4$ &  $-0.4\pm0.7$ & $\phantom{0}8.21\pm0.19$    \\
     \kpitag &   $13.77\pm0.10$ &   $23.5\pm0.5$ &  $-2.4\pm0.8$ & $\phantom{0}3.87\pm0.14$    \\
    \piontag &   $14.38\pm0.10$ &   $33.0\pm0.5$ &  $\phantom{-}5.2\pm0.8$ &  $\phantom{0}1.67\pm0.10$   \\
   \othertag &   $\phantom{0}9.61\pm0.08$  &   $41.9\pm0.6$ &  $\phantom{-}4.6\pm0.9$ &  $\phantom{0}0.25\pm0.04$   \\\hline

         All &   $74.60\pm0.12$ &           &          & $30.4\pm0.3$   \\\hline\hline
\end{tabular}
%\vspace{0.5cm}
\end{center}
\end{table}

%--------------------------------------
% Figure : mES, DeltaE distributions. 
%--------------------------------------
\begin{figure}[!htb]%1
\begin{center}
\vspace{1.0cm}
\scalebox{0.6}{\includegraphics{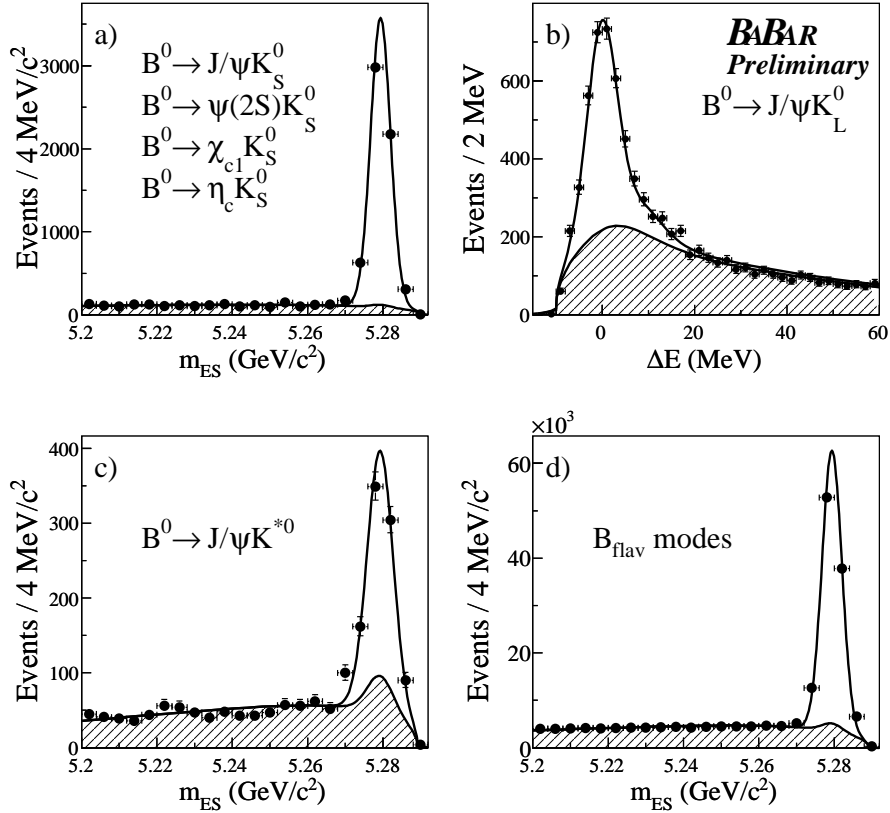}}
\caption{
Distributions for $B_{\CP}$ and \bflav\ candidates satisfying the tagging and vertexing requirements:
a) \mes\ for the final states $J/\psi\KS $, $\psi(2S)\KS$, $\chi_{c1}\KS$, and $\etac\KS$, 
b) $\Delta E$ for the final state $\jpsi\KL$,
c) \mes\ for $J/\psi K^{*0}(K^{*0}\to \KS\pi^0)$, and 
d) \mes\ for the \bflav\ sample. In each plot, the shaded region is the estimated background contribution.}
\label{fig:bcpsample}
\end{center}
\end{figure}
%--------------------------------------
%------------------------------------
% Table : sin(2\beta) results table
%------------------------------------
\begin{table}[!htb] 
\begin{center}
\caption{
Number of events $N_{\rm tag}$ in the signal region after tagging and vertexing requirements, 
signal purity $P$ including the contribution from peaking background,
and results of fitting for \CP\ asymmetries in the $B_{\CP}$ sample and various subsamples.
In addition, results on the $B_{\rm flav}$ and charged $B$ control samples
test that no artificial \CP\ asymmetry is found where we expect no \CP\ violation ($\stwob=0$).
Errors are statistical only. The  signal region is $5.27 < \mes < 5.29 \gevcc$
 ($\vert \Delta E \vert < 10\mev$ for $\jpsi \KL$).\label{tab:result}}
\vspace{0.5cm}
\begin{tabular}{lrcc}\hline\hline
 Sample  & $N_{\rm tag}$ & $P(\%)$ & \multicolumn{1}{c}{$\ \ \ \stwob$}
\\ 
 Full \CP\ sample                                   & $11496$        & $76$
 &  $0.710\pm0.034$   \\
\hline
$\jpsi\KS$,$\psitwos\KS$,$\chicone\KS$,$\etac\KS$   & $6028$        & $92$
 &  $0.713\pm0.038$   \\
$\jpsi \KL$                                         & $4323$        & $55$
 &  $0.716\pm0.080$   \\
$\jpsi\Kstarz (\Kstarz \to \KS\piz)$                 & $965$         & $68$
  &  $0.526\pm0.284$   \\
\hline
1999-2002 data                                       &  $3084$        &  $79$
      &  $0.755\pm0.067$   \\
2003-2004 data                                       &  $4850$        &  $77$
      &  $0.724\pm0.052$   \\
2005-2006 data                                       &  $3562$        &  $74$
      &  $0.663\pm0.062$   \\
\hline
\hline
\multicolumn{1}{l}{$\jpsi\KS$, $\psitwos\KS$, $\chicone\KS$, $\etac\KS$ only  $(\eta_f=-1)$ }  \\
\hline
$\ \ \jpsi \KS$ ($\KS \to \pi^+ \pi^-$)    & $4076$        & $96$       &  $0.715\pm0.044$ \\
$\ \ \jpsi \KS$ ($\KS \to \pi^0 \pi^0$)    & $988$        & $88$       &  $0.581\pm0.105$\\
$\ \ \psi(2S) \KS$ ($\KS \to \pi^+ \pi^-$) & $622$        & $83$       &  $0.892\pm0.120$\\
$\ \ \chicone \KS $                        & $279$         & $89$       & $0.709\pm0.174$ \\
$\ \ \etac\KS $                            & $243$        & $75$       &  $0.717\pm0.229$\\
\hline
$\ $ \leptontag\ category                & $703$        & $97$       &  $0.754\pm0.068$\\
$\ $ \kaonitag\ category                 & $900$        & $93$       &  $0.713\pm0.066$   \\
$\ $ \kaoniitag\ category                & $1437$        & $91$       & $0.711\pm0.075$ \\
$\ $ \kpitag\ category                   & $1107$        & $89$       & $0.635\pm0.117$   \\
$\ $ \piontag\ category                  & $1238$        & $91$       & $0.587\pm0.175$  \\
$\ $ \othertag\ category                 & $823$        & $89$       &  $0.454\pm0.469$   \\
\hline\hline
$B_{\rm flav}$ sample                    & $112878$      & $83$       &  $0.016\pm0.011$   \\
\hline
$B^+$ sample                             & $27775$      & $93$       &  $0.008\pm0.017$   
\\\hline\hline
\end{tabular}
\end{center}
%\vspace{1.0cm}
\end{table}
%--------------------------------------
With the exception of the $\jpsi\KL$ mode, we use the beam-energy substituted mass 
$\mes=\sqrt{{(E^{*}_{\rm beam})^2}-(p^{*}_B)^2}$
to determine the composition of our final sample, where $E^{*}_{\rm beam}$ and $p_B^{*}$ are the 
beam energy and $B$ momentum in the $\epem$ center-of-mass frame.
For the $\jpsi\KL$ mode we use the difference $\Delta E$ between the candidate 
center-of-mass energy and $E^{*}_{\rm beam}$. 
The composition of our final sample is shown in Fig.~\ref{fig:bcpsample}.
We use events with $\mes > 5.2 \gevcc$ ($\Delta E < 80\mev$ for $\jpsi \KL$) 
in order to determine the properties of the background contributions.
We define a signal region $5.27 < \mes < 5.29 \gevcc$  
($\vert \Delta E \vert < 10\mev$ for $\jpsi \KL$)
that contains {\bf{$ $}} \CP\ candidate events that satisfy the tagging and vertexing 
requirements as listed in Table~\ref{tab:result}.

For all modes except $\eta_c \KS$ 
and $\jpsi\KL$ we use simulated events to estimate the fractions of
events that peak in the $\mes$ signal region
due to cross-feed from other decay modes (Peaking background).
For the $\eta_c\KS$ mode the cross-feed fraction is determined from a fit
to the $m_{KK\pi}$ and \mes\ distributions in data.
For the $\jpsi\KL$ decay mode, the sample composition, 
effective $\eta_f$, and
$\Delta E$ distribution of the individual background sources are
determined either from simulation (for $B\to\jpsi X$) 
or from the $m_{\ell^+ \ell^-}$ sidebands in data (for fake $\jpsi\to \ell^+ \ell^-$).

We determine \stwob with a simultaneous maximum likelihood fit to the 
\deltat\ distribution of the tagged $B_{\CP}$ and \bflav\ samples. The 
\deltat\ distributions of the $B_{\CP}$ sample are modeled  
by Eq.~\ref{eq:timedist} with $|\lambda|=1$.
Those of the \bflav\ sample evolve according to the known frequency for 
flavor oscillation in $B^0$ mesons. We assume that the observed amplitudes for the \CP 
asymmetry in the $B_{\CP}$ sample and for flavor oscillation in the \bflav\ sample
are reduced by the same factor $1-2\mistag$ due to flavor mistags.
The \deltat distributions for the signal are convolved with a 
resolution function common to both the \bflav and $B_{\CP}$ samples, 
modeled by the sum of three Gaussians~\cite{ref:bigprd}.
Backgrounds are incorporated with an empirical
description of their \deltat spectra, containing zero and 
non-zero lifetime components convolved with a resolution
function~\cite{ref:bigprd} distinct from that of the signal.

There are 65 free parameters in the fit: \stwob (1),
the average mistag fractions $\mistag$ and the
differences $\Delta\mistag$ between \Bz\ and \Bzb\ mistag fractions for each
tagging category (12), parameters for the signal \deltat resolution (7),
parameters for \CP\ background time dependence (8), and the difference between
$\Bz$ and $\Bzb$ reconstruction and tagging efficiencies (7); for 
\bflav\ background, time dependence (3), \deltat resolution
(3), and mistag fractions (24). For the \CP\ modes (except for $\jpsi \KL$), 
the apparent \CP\ asymmetry of the non-peaking background in each tagging
category is allowed to be a free parameter in the fit.  

We fix $\tau_{\Bz}=1.530\,\ps$, $\deltamd
=0.507\,\ps^{-1}$~\cite{ref:pdg2006}, $\vert \lambda \vert = 1$, and $\Delta \Gamma_d=0$.
The determination of the mistag fractions and \deltat resolution
function parameters for the signal is dominated by the large \bflav\ sample. 
We determine background parameters mainly from events 
outside the peaks in the $m_{\rm ES}$ and $\Delta E$ distribtions, as shown 
in Fig.~\ref{fig:bcpsample}.

\begin{figure}[!htb]
\begin{center}
\vspace{1.0cm}
\begin{center}
\scalebox{0.5}{\includegraphics{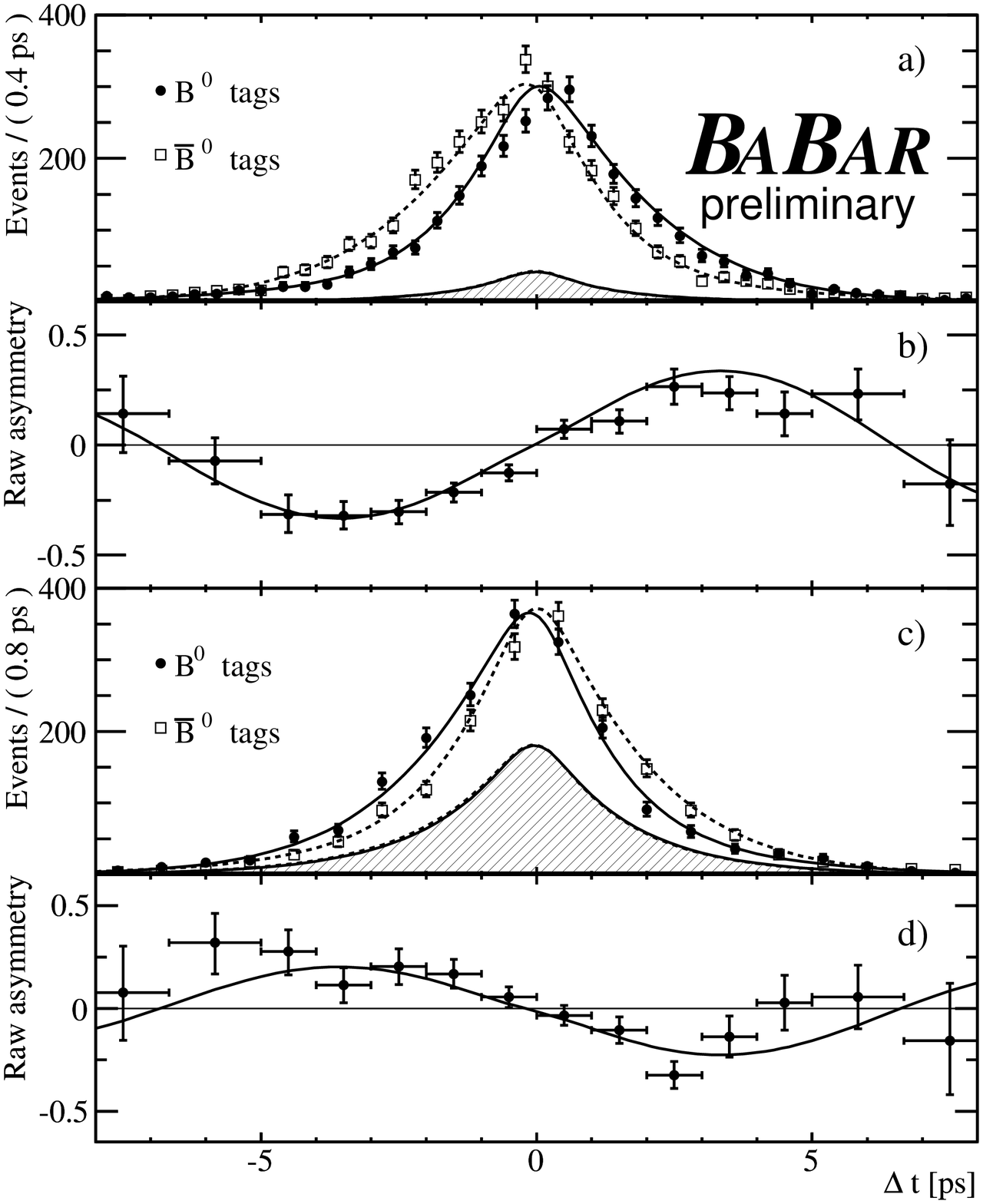}}
\end{center}
\vspace{1.0cm}
\caption{
a) Number of $\eta_f=-1$ candidates ($J/\psi \KS$, $\psi(2S) \KS$, $\chicone \KS$, and $\eta_c \KS$)
in the signal region with a \Bz tag ($N_{\Bz }$) and with a \Bzb tag ($N_{\Bzb}$), and 
b) the raw asymmetry $(N_{\Bz}-N_{\Bzb})/(N_{\Bz}+N_{\Bzb})$, as functions of \deltat.
Figures c) and d) are the corresponding distributions for the $\eta_f=+1$ mode $J/\psi \KL$.  
All distributions exclude \othertag-tagged events. The solid (dashed) curves 
represent the fit projections
in \deltat for \Bz (\Bzb) tags. The shaded regions represent the estimated background contributions.
\label{fig:cpdeltat}}
\end{center}
\end{figure}

The fit to the $B_{\CP}$ and \bflav\ samples yields
\begin{eqnarray}
\stwob = \fitstwob \pm \statstwob \stat \pm \syststwob\syst\nonumber.
\end{eqnarray}
\noindent
Figure~\ref{fig:cpdeltat} shows the \deltat distributions and 
asymmetries in yields between events with \Bz tags and \Bzb tags for the
$\eta_f=-1$ and $\eta_f = +1$ samples as a function of \deltat,
overlaid with the projection of the likelihood fit result.
%--------------------------------------
% Figure : Deltat distributions etc ...
%--------------------------------------
%--------------------------------------
We perform a separate fit with only the cleanest $\eta_f=-1$ sample, in which we treat
both $\vert\lambda\vert$ and $\rm{sin}2\beta$ as free parameters. 
We do not use the modes $J/\psi K^{*0}$ and $J/\psi \KL$
to minimize the  dependence of the results on the background parametrization.
We obtain  $|\lambda| = \fitlambda \pm \statlambda \stat \pm \systlambda \syst$.
The correlation between the coefficients multiplying the $\sin(\deltamd \deltat)$
and $\cos(\deltamd \deltat)$ terms in Eq.~\ref{eq:timedist} is $-1.2\%$.

%------------------------------
\section{SYSTEMATIC UNCERTAINTIES}
\label{sec:Systematics}
%------------------------------
The systematic uncertainties on \stwob and $|\lambda|$ 
are summarized in Table~\ref{tab:systematics}.
These include the uncertainties in the level and \CP\ asymmetry of
the peaking background, the assumed parameterization of the \deltat\ resolution function,
possible differences between the \bflav\ and $B_{CP}$ tagging performances 
and \deltat\ resolution functions, 
knowledge of the event-by-event beam spot position,
and the possible interference between the suppressed $\bar b\to \bar u c \bar d$ amplitude with
the favored $b\to c \bar u d$ amplitude for some tag-side 
$B$ decays~\cite{ref:dcsd}. 
In addition, we include the variation due
to the assumed values of $\Delta m_d$ and $\tau_B$~\cite{ref:pdg2006}.
We also assign the change in the measured \stwob\ as the corresponding 
systematic uncertainties when we let $\vert \lambda \vert$
to be a free parameter in the fit and when we set $\Delta \Gamma_d / \Gamma_d = \pm 0.02$, 
the latter being
considerably larger than SM estimates~\cite{ref:dg}. 
The total systematic error on \stwob\ ($\vert \lambda \vert$)
is $0.019$ ($0.017$).
%---------------------------------
% Table : Systematic Error Table
%---------------------------------
\begin{table}[!ht]
\vspace{0.3cm}
\begin{center}
\caption{
Systematic uncertainties on \stwob and $\vert \lambda \vert$.\label{tab:systematics}}
\vspace{0.3cm}
\begin{tabular}{lcc}\hline\hline
 Source                                                               & $\sigma(\stwob)$ & $\sigma(\vert \lambda \vert)$\\ \hline
\CP\ backgrounds                                                      & 0.007            & 0.002\\
\deltat\ resolution function                                          & 0.008            & 0.002\\
$\jpsi \KL$ backgrounds                                               & 0.007            & N/A \\
Mistag fraction differences                                           & 0.009            & 0.007\\
Beam spot                                                             & 0.008            & 0.004\\
$\deltamd$, $\tau_B$, $\Delta \Gamma_d / \Gamma_d$, $\vert \lambda \vert$ & 0.003            & 0.001\\
Tag-side interference                                                 & 0.002            &  0.014\\
MC statistics                                                         & 0.003            & 0.005\\ \hline
Total systematic error                                                & 0.019            & 0.017 \\\hline\hline
\end{tabular}
\vspace{0.5cm}
\end{center}
\end{table}
%--------------------------------------

The large $B_{\CP}$ sample allows a number of consistency
checks, including separation of the data by decay mode and tagging category.
The results of those checks are listed in Table~\ref{tab:result}. 
We observe no statistically significant asymmetry
from fits to the control samples of non-\CP decay modes.

%------------------------------
\section{SUMMARY}
\label{sec:Summary}
%------------------------------
In summary, we report on improved measurements of $\stwob$ 
and $|\lambda|$ that 
supersede our previous result~\cite{ref:babar2004}.
We measure $\stwob=\fitstwob\pm\statstwob\stat\pm \syststwob\syst$ 
and $|\lambda|=\fitlambda\pm\statlambda\stat\pm\systlambda\syst$.
The updated value of $\stwob$ is consistent with the current 
world average~\cite{ref:HFAG2006} and the theoretical estimates of the 
magnitudes of CKM matrix elements in the context of the 
SM~\cite{ref:CKMconstraints}. 
The theoretical uncertainty on the interpretation of the measurement of \stwob\ in 
these modes is approximately 0.01~\cite{Ciuchini:2005mg}. 

%------------------------------
\section{ACKNOWLEDGMENTS}
\label{sec:Acknowledgments}
%------------------------------
We are grateful for the 
extraordinary contributions of our \pep2\ colleagues in
achieving the excellent luminosity and machine conditions
that have made this work possible.
The success of this project also relies critically on the 
expertise and dedication of the computing organizations that 
support \babar.
The collaborating institutions wish to thank 
SLAC for its support and the kind hospitality extended to them. 
This work is supported by the
US Department of Energy
and National Science Foundation, the
Natural Sciences and Engineering Research Council (Canada),
Institute of High Energy Physics (China), the
Commissariat \`a l'Energie Atomique and
Institut National de Physique Nucl\'eaire et de Physique des Particules
(France), the
Bundesministerium f\"ur Bildung und Forschung and
Deutsche Forschungsgemeinschaft
(Germany), the
Istituto Nazionale di Fisica Nucleare (Italy),
the Foundation for Fundamental Research on Matter (The Netherlands),
the Research Council of Norway, the
Ministry of Science and Technology of the Russian Federation, and the
Particle Physics and Astronomy Research Council (United Kingdom). 
Individuals have received support from 
the Marie-Curie IEF program (European Union) and
the A. P. Sloan Foundation.

%------------------------------
% Bibiliography
%------------------------------


\begin{thebibliography}{99}
\bibitem{ref:chargeconj}
Charge-conjugate reactions are included implicitly unless otherwise specified.
\end{thebibliography}
\end{document}